\documentstyle[aps,twocolumn,prb,epsf]{revtex}

\begin{document}

\newcommand{\be}{\begin{equation}}
\newcommand{\ee}{\end{equation}}
\newcommand{\bn}{\begin{eqnarray}}
\newcommand{\en}{\end{eqnarray}}
\def\beq{\begin{eqnarray}}
\def\eeq{\end{eqnarray}}
\def\lsim{\:\raisebox{-0.5ex}{$\stackrel{\textstyle<}{\sim}$}\:}
\def\gsim{\:\raisebox{-0.5ex}{$\stackrel{\textstyle>}{\sim}$}\:}
\def\d{\displaystyle}
\def\u{\underbar}

\draft

\twocolumn[\hsize\textwidth\columnwidth\hsize\csname @twocolumnfalse\endcsname

\title{ Orbital selective insulator-metal transition in 
${\rm V_{2}O_{3}}$ under external pressure}

\author{M. S. Laad$^1$, L. Craco$^2$ and E. M\"uller-Hartmann$^2$}

\address{$^1$Department of Physics, Loughborough University, LE11 3TU, UK \\
$^2$Institut f\"ur Theoretische Physik, Universit\"at zu K\"oln, 
77 Z\"ulpicher Strasse, D-50937 K\"oln, Germany}

\date{\today}
\maketitle

\widetext

\begin{abstract}
We present a detailed account of the physics of Vanadium sesquioxide 
(${\rm V_{2}O_{3}}$), a benchmark system for studying correlation induced 
metal-insulator transition(s).  Based on a detailed perusal of a wide 
range of experimental data, we stress the importance of multi-orbital 
Coulomb interactions in concert with first-principles LDA bandstructure 
for a consistent understanding of the PI-PM MIT under pressure. Using 
LDA+DMFT, we show how the MIT is of the orbital selective type, driven 
by large changes in dynamical spectral weight in response to small 
changes in trigonal field splitting under pressure. Very good quantitative 
agreement with ($i$) the switch of orbital occupation and ($ii$) $S=1$ 
at each $V^{3+}$ site across the MIT, and ($iii$) carrier effective mass 
in the PM phase, is obtained.  Finally, using the LDA+DMFT solution, we 
have estimated screening induced renormalisation of the local, 
multi-orbital Coulomb interactions.  Computation of the one-particle 
spectral function using these screened values is shown to be in excellent 
quantitative agreement with very recent experimental (PES and XAS) results.   
These findings provide strong support for an orbital-selective Mott transition
in paramagnetic ${\rm V_{2}O_{3}}$.    
\end{abstract}

\pacs{PACS numbers: 71.28+d,71.30+h,72.10-d}

]

\narrowtext

\section{INTRODUCTION}
  
  Correlation driven metal insulator transitions have remained unsolved 
problems of solid state theory of electrons in solids for more than five 
decades. The pioneering work of Mott,~\cite{[1]} and of Gutzwiller,~\cite{[2]}
Kanamori,~\cite{[3]} and Hubbard~\cite{[4]} involved a detailed exposition of
the view suggesting that description of such phenomena lay outside the 
framework of band theory.  Subsequent, more recent developments, including 
discovery of high-$T_{c}$ superconductors, rare-earth based systems, colossal
magnetoresistive oxides along with whole families of other systems have 
shown that strong electronic correlations give rise to widely unanticipated,
fundamentally new types of metallic behaviors, namely, non-Fermi liquid 
metals. A host of very careful studies now clearly show that these anomalous 
responses seem to be correlated with the existence of a correlated metallic
state on the border of a Mott insulator in $d$-band oxides,~\cite{[5]} or to
a metallic state in proximity to a localisation-delocalisation transition of
$f$-electrons in rare-earth compounds.~\cite{[6]} 

The corundum lattice-based transition metal oxide system vanadium sesquioxide 
(${\rm V_{2}O_{3}}$) has been of interest for more than five decades 
as a classic (and by now a textbook) case of an electronic system 
with $S=1/2$ local moments at each site exhibiting the phenomenon of the 
correlation-driven Mott-Hubbard metal-insulator transition.~\cite{[1],[4]}  
Widely accepted wisdom has it that this is one of the few cases where 
modelling in terms of a simple one-band Hubbard model is appropriate.  
Over the last few years, experimental and theoretical work has forced a 
revision of this view, leading on the one hand to a spurt of new approaches, 
and on the other to an expansion of our perception of what is new and 
important in the physics of TM-oxides in general. Specifically, 
taken by themselves as well as in combination with properties of other systems
like ruthenates, CMR manganites, etc, these new studies force one to refocus
the attention in terms of the strong coupling and interplay of spin and 
orbital degrees of freedom (not to be confused with the usual spin-orbit 
coupling, though that may also be relevant in some situations) and of their
combined influence on the nature of charge and spin dynamics in TM oxides.

In what follows, we aim to present a far from complete view of the questions
which are posed by these new studies.  Focusing our attention to the early 
vanadium oxides, ${\rm V_{2}O_{3}}$, we will start with a somewhat detailed 
perusal of earlier work, review key recent experiments, and follow 
them up with a discussion of their implications for theory.  Finally, we 
will suggest a rather detailed scenario  for correlation-induced metal 
insulator transitions in ${\rm V_{2}O_{3}}$ that ties together essential 
experimental constraints in one picture.  In doing so, we will give a 
detailed description of our theoretical modelling using a combination of 
the local density approximation (LDA) combined with multi-orbital dynamical 
mean field theory (DMFT) using the iterated perturbation theory (IPT) as the 
``impurity'' solver.

In the first part, we will confine ourselves to summarizing known -and not 
so well-known experimental results on the effect of external pressure and 
${\rm Cr}$ doping on the thermodynamic and transport properties of 
${\rm V_{2}O_{3}}$ along with the magnetic and orbital structure (and 
their changes) across the metal insulator transition.  

In the second part, we will first review the earlier theories for the MIT
in terms of the one-band Hubbard model as well as the more recent multiband
Hubbard model (which allows a description in terms of a $S=1$ model).  
Finally, we will propose a new scenario: one where the abrupt change in the 
character of spin (and presumably also orbital) correlations across the MIT 
is described within the strong correlation scenario.  In particular, we will 
show how a two-fluid (i.e., orbitally selective) description can be derived 
from first principles, and demonstrate how the properties of 
${\rm V_{2}O_{3}}$ can be understood in this scenario. 

\section{EXPERIMENTAL REVIEW}

The magnetic structure of ${\rm V_{2}O_{3}}$ has been measured a long time 
ago.~\cite{[7]} Its interpretation has however remained a subject of 
controversy. The robust aspects are: the antiferro-insulator
(AFI) is characterized by AF order 
which spontaneously breaks the crystal symmetry of the corundum lattice 
-and in modern parlance, it corresponds to the C-type AF order~\cite{[8]} 
with one F and two AF bonds in the hexagonal plane.  The vertical V-V 
pairs (with ${\rm V^{3+}}$) form dimers in the solid; these are aligned 
antiferromagnetically with the inplane V-V pairs, and ferromagnetically 
with other vertical V-V pairs.  In terms of these dimers, the corundum 
lattice can be viewed as a distorted simple cubic lattice. Based on the 
$S=1/2$ picture of Castellani {\it et al.}, the spin waves were
``characterized'' in terms of a Heisenberg-like model.~\cite{[8]}  This 
was a commonly accepted picture for almost two decades, until recent 
experimental results forced one to reanalyze it.  These are:

$(i)$ Change of magnetic correlations across the AFI to AF-metal (AFM) phase 
transition in ${\rm V_{2-y}O_{3}}$.  In a one-band Hubbard model scenario, 
one would expect that the magnetic correlations in the AFM should be 
a remnant of those in the AFI, for e.g, broadened spin-waves of the AFI.  
However, measurements revealed, surprisingly,~\cite{[7]} that the transport 
and thermodynamics is due to onset of magnetic order which seems to 
be totally unrelated to that in the AFI. The AFM is characterized 
by incommensurate order with ${\bf Q}||c$, in contrast to that for the 
AFI, characterized by ${\bf Q}=(1/2,1/2,0)$ (hexagonal notation).  This 
order in the metal is reminiscent of a small-moment SDW derived from a 
Fermi surface instability.~\cite{[7]} Constant energy scans in INS also 
show that the magnetic fluctuations are far from conventional spin-waves; 
they are more reminiscent of particle-hole modes with extremely small 
correlation length of about 14~\AA, much smaller than that characterizing 
the spin-waves in the AFI. Finally, the AFI-PI transition (P meaning 
paramagnetic) occuring around 
$T_{c}=150$K is second order, once again accompanied by an abrupt change 
in short-ranged spin correlations. An interesting correlation concerns the 
change of lattice structure from monoclinic (AFI) to corundum (PI,PM), 
and may be linked to a change in orbital correlations.

Further, abrupt jump of the crystal volume (without change of symmetry) is 
also observed across the PI-PM transition at higher $T$.~\cite{[9]}
The $c$-axis distance decreases abruptly at the PI-PM transition, while 
the $a$-axis distance slightly increases, and this change needs to be 
correlated with the conductivity jump at the transition (see below).  
                                                            
$\rightarrow$ This abrupt switching of the spin correlations across the 
AFI-AFM transition is inexplicable within an effective correlated one-band 
scenario. The AF-P transition(s) cannot be considered as order-disorder 
transitions of the usual type.

$\rightarrow$ Are orbital degrees of freedom involved in the monoclinic to
corundum structural change at the MIT?

$\rightarrow$ What is the specific relation between the lattice contraction 
and the jump in the $dc$ conductivity at the MIT?  Does it involve changes in 
carrier concentration, or in the carrier mobility?

$(ii)$ Recent X-ray experiments by Park {\it et al.}~\cite{[10]} have 
revealed the existence of an admixture of $(e_{g1}^\pi,e_{g2}^\pi)$ and 
$(e_{gi}^\pi,a_{1g})$ with $i=1,2$ and a spin $S=1$ on each V site, in 
contrast to the $S=1/2$ proposed in Ref.~\onlinecite{[8]}. In addition, large 
differences in their ratio have been found in the AFI, PI and PM phases.  
In particular, this ratio is $(e_{g1}^\pi,e_{g2}^\pi):(e_{gi}^\pi,a_{1g})$ 
is $2:1$ in the AFI, $1.5:1$ in the PI and $1:1$ in the PM phase.

The AFI phase is also characterized by a monoclinic distortion involving a 
uniform rotation of all the V-V pairs,~\cite{[11]} an observation which 
puts specific constraints on possible orbital order in the AFI.  However, 
the corundum structure is recovered in the PI phase. (Does this change 
involve switching of the orbital correlations across AFI-PI boundary, and 
if so, how?) Based on this, Ezhov {\it et al.}~\cite{[27]} have proposed a 
$S=1$ model without orbital degeneracy as an alternative starting point 
to Ref.~\onlinecite{[8]}. More studies along this line have been done by 
Tanaka.~\cite{[11]} An alternative point of view by Shiina
{\it et al.}~\cite{[12]} pictures the AFI as a C-type AF ordered 
state with $S=2$ and ferro-orbital ordering.  

$\rightarrow$ Starting point should involve $S=1$, suggesting use of 
multiband models is necessary to describe ${\rm V_{2}O_{3}}$.  In orbitally 
degenerate cases, the ground state is simultaneously spin and orbital 
ordered, and the strong coupling between the elementary excitations 
involving both spin and orbital flips is expected to result in emergence 
of qualitatively new behavior (not, however, if the scenario of 
Ref.~\onlinecite{[11]} is taken to be valid). It is important to notice that 
even in the para-orbital/magnetic state(s), additional strong scattering 
resulting from coupled spin and orbital fluctuations might cause pronounced 
deviations from expectations based on (correlated) Fermi liquid theory.

$\rightarrow$  Notice that this change of orbital occupation also implies an
important role for the trigonal distortion (this would act like an external 
field in orbital pseudospin space).  This quantity determines the occupation 
of the relevant orbitals, and in a coupled spin-orbital system, determines 
the effective exchange interactions (and hence the magnetic structure).
In fact, the importance of this quantity, and its pressure (strain) 
dependence, has been identified in many members of the corundum based 
oxides~\cite{[13]} (${\rm Fe_{2}O_{3}}$ in connection with the Morin or 
the strain-induced spin-flop transition, ${\rm Cr_{2}O_{3}}$, in the same 
connection, ${\rm Ti_{2}O_{3}}$ as manifested by an anomaly in the 
$T$-dependence of the $A_{1g}$ phonon frequency across the MIT), suggesting 
a related common origin of the varied manifestations observed in this 
structurally related class of oxides.

$\rightarrow$  It might be interesting to look at possible anomalies of the
$A_{1g}$-mode frequency in Raman scattering measurements across the P-MIT in
${\rm V_{2}O_{3}}$.

$(iii)$ Resonant X-ray scattering (RXS) measurements at the V K-edge have 
been performed by Paolasini {\it et al}.~\cite{[14]}  The resonant Bragg 
peaks observed in the AFI phase are interpreted in terms of a $3d$ orbital 
ordering. However, the authors of Ref.~\onlinecite{[14]} interpreted their 
results within the picture of Castellani {\it et al.}, leading to 
difficulties with $(i)$ and $(ii)$.

$\rightarrow$ One requires a re-interpretation in a way consistent with 
$(i)$ and $(ii)$, for e.g, see Mila {\it et al.} An Aside: Is the 
Kugel-Khomskii type of modelling required for the AFI?

$(iv)$  Finally, Lovesey {\it et al.}~\cite{[16]} argue that the resonant 
Bragg peaks arise due to ordered orbital magnetic moments of the $V$ ions.  
Indeed, a large orbital contribution $M_{L}/M_{S} \simeq -0.3$ to the total 
magnetic moment was claimed in Ref.~\onlinecite{[16]}. If correct, the 
effect of spin-orbit coupling might become important, as argued in 
Ref.~\onlinecite{[11]}. On the other hand, if the spin-orbit coupling 
leads to small effects on the electronic structure, a re-interpretation 
would be in order.

$\rightarrow$ How important is the role of spin-orbit coupling in the AFI 
phase?

$(v)$ Optical spectroscopy provides a detailed picture of charge dynamics.  
Careful studies by Thomas {\it et al.}~\cite{[17]} reveal all the 
characteristics of a strongly correlated system: an ''upper Hubbard band'' 
(UHB) feature with a threshold in the AFI (and PI), and a sharp, 
quasicoherent feature, along with an intense mid-infra-red peak and the 
remnant of the UHB on the PM side. Calculations within the framework of 
a $S=1/2$, one-band Hubbard model~\cite{[18]} claim to obtain very good 
agreement with the observed spectra.  However, to achieve this, the 
Hubbard $U$ has to be changed by a factor of 2 on going from the PI to 
the PM state, which is hardly conceivable. Alternatively, the effective 
hopping, or the degree of itinerance, should increase in the metallic 
state, leading to a change in the effective $U/t$ value, and, as is 
ubiquitous in strongly correlated scenarios, to a large transfer of 
spectral weight. In particular, given $(i)-(iii)$, this should involve 
carriers coupled to spin-orbital degrees of freedom along with the 
concomitant lattice distortion.

$\rightarrow$  What is the specific nature of the correlation between the 
change in spin-orbital correlations across the MIT and the tendency to 
increased itinerance which drives this transition?  In particular, is the 
abrupt change in the $e_{g1}^\pi e_{g2}^\pi:e_{gi}^\pi a_{1g}$ ratio 
related to increase 
in carrier concentration or the carrier mobility (kinetic energy)?
Notice that (see below) the dc transport data can be reconciled with 
increase in the carrier {\it density},~\cite{[19]} so this is an important 
point deserving more attention.  Obviously, to make a plausible correlation 
between the two requires experimental characterization, as well as a proper 
treatment of these coupled correlations.

$(vi)$ Photoemission Spectroscopy (PES)~\cite{[20]} reveals further proof 
of the correlation-driven character of the insulator-metal transition. At 
high $T$, the data is claimed to be consistent with a ``thermally smeared 
quasicoherent'' peak, something within reach of single-site theories.  
However, it is more conceivable that strong inelastic scattering from 
coupled spin-orbital excitations gives rise to non-quasiparticle dynamics 
in the PM phase. At lower $T$, appreciable changes are observed in the PES 
spectra across the AFI/PM and PI/PM transitions, with the characteristic 
transfer of spectral weight from high to low energy over a scale of almost 
4~eV (notice that $T_{MI}\simeq 300$K) implying a drastic rearrangement of 
electronic states over a wide energy scale.~\cite{[21]} In the PM state, 
the PES spectrum shows the asymmetric two-peak structure with a Fermi edge 
(but we draw attention to the fact that this low-energy ''peak'' is 
anomalously broad, suggesting non-quasiparticle dynamics), 
while a clear opening of a spectral gap ($E_{g}$) occurs in the AFI and PI 
phases.  In the AFI phase, $E_{g}^{AFI}\simeq 0.3$eV, while for the PI, 
$E_{g}^{PI}\simeq 0.23$eV, with a more symmetric lineshape.

Earlier PES studies across the PI-PM transition have been controversial; in
particular, the question of the $T$-dependence of the low energy spectral 
weight was not settled till recently.  Quite recently, this question has been
answered by the Michigan group,~\cite{Mo-2004} and the $T$-dependent 
renormalized, and heavily damped ``quasiparticle'' contribution has 
indeed been observed.   

Details of the PES spectra at high $T$ are seemingly well captured by a 
DMFT applied to a multiband Hubbard model in combination with the actual 
LDA bandstructure (see below for more details).~\cite{[21]}  There are 
still some discrepancies between LDA+DMFT and experiment at lower 
$T < 400~K$, however: the ``quasiparticle peak'' is too broad by a factor 
of $2-3$, and the details of the PES lineshape in the PI still remain to 
be calculated.  It is possible that screening-induced renormalization of 
$U$, etc. needs to be included; however, it is a very difficult task to do this 
from an {\it ab initio} starting point.  Alternatively, or in concert with 
the above, the dynamical effect of intersite correlations might be expected 
to become increasingly important at lower $T$.  Such effects are out of scope 
of LDA+DMFT, and require extensions to treat dynamical effects of spatial
correlations, a more demanding task.

Given this, the interpretation of the PES spectrum will also need a
re-examination.  In particular, the possible importance of short-ranged 
spin-orbital correlations might be required to understand the anomalously 
broad, low-energy feature observed in PES in the PM phase. 

In a coupled spin-orbital system, the degree of itinerance is
directly related to the changes in spin and orbital correlations coupled to 
possible structural changes. In course of its hopping motion, an electron is
scattered by coupled spin-orbital excitations, i.e, by simultaneous flipping
of spin and orbital pseudospins.  In the AFI, this is not enough to destroy
AF/O order (O meaning orbital). An understanding of the change in AF/O 
correlations across the MIT is necessary to understand the enhancement of 
itinerance.  In particular, within the framework of Shiina {\it et al.}, 
do the changes involve a cooperative melting of the AF/O order of the AFI?  
How does one then try to understand the PI/PM transition?

$(vii)$ One of the most spectacular hallmarks of the I-M transition in 
${\rm V_{2}O_{3}}$ is the sharp jump in conductivity by seven orders of 
magnitude! Is the jump of $\sigma(T)$~\cite{Limel} driven by a jump in 
the carrier density at the 
transition, or by an increase in the mobility?~\cite{[19]}  Hall effect 
measurements would be a probe to answer this question (complications due to 
possible relevance of spin-orbit coupling, if important).  On the barely 
metallic (close to the AFI) side (${\rm V_{2-y}O_{3}}$), the Hall constant 
$R_{H}(T)$ shows behavior reminiscent of the cuprates;~\cite{[22]} it is
strongly $T$-dependent, increasing with decreasing $T$ with a peak around 
the AF ordering temperature, followed by a drop at lower $T$.
The $T$-dependence gets weaker with increasing metallicity ($y$).  More 
similarities with the normal state of the high-$T_{c}$ cuprates are seen 
in the different $T$ dependences of $\rho(T) \simeq T^{3/2}$ and 
$\cot \theta_{H}(T) \simeq aT^{2}+b$ for small $y$, which evolves into 
more conventional FL behavior with increasing $y$.~\cite{[22]}  Such a 
behavior would mandate strong local moment scattering in the metallic phase.  
Given the strong correlation signatures observed globally, a description 
in terms of vagaries of the Fermi surface is untenable.

$\rightarrow$ Does the I-M transition involve a jump in the carrier density?

$\rightarrow$ How does one understand the anomalous features of the Hall data
near the MIT?  In particular, similarity to cuprates suggests that such 
anomalies might be more general manifestations of the breakdown of 
Fermi-liquid theory (FLT) near 
the Mott transition to a Mott-Hubbard antiferromagnet, as opposed to a Slater
antiferromagnet. The observation of overdamped spin waves with extremely 
short correlation length and anomalously broad linewidth is also reconcilable 
in terms of a strong scattering scenario.

\subsection{Summary of experimental results}

In conclusion, experimental results reveal very interesting points concerning
the nature of the ground states and collective excitations in the different
phases of ${\rm V_{2}O_{3}}$.  

\subsubsection{In the AFI phase}  

$(i)$ C-type AF order with ferro-type concomitant orbital order.  In terms of
the V-V pairs, it corresponds to C-type AF order on a distorted simple cubic
lattice.  A Kugel-Khomskii type of model is required to derive the AF/FO order.
The picture requires consistency with $S=1$ at each $V$ site, and with a 
mixture of $(e_{g}^\pi,e_{g}^\pi)$ and $(e_{gi}^\pi,a_{1g})$ on each 
V-V pair.

$(ii)$ Spin wave spectra in the insulator should be consistent with exchange 
constants ($J_{ij}^{ab}$) set by the FO order (FO order is consistent with the 
monoclinic distortion involving uniform rotation of {\it all} V-V pairs in 
the AFI).  Also (see Mila {\it et al.}), it can be reconciled with anomalous
X-ray scattering results.

\subsubsection{In the PM phase}

$(i)$ no AF order (not even a remnant of AF-LRO of the AFI).  I-M transition 
strongly first order.  A jump in the 
$(e_{g1}^\pi,e_{g2}^\pi):(e_{gi}^\pi,a_{1g})$ from
$2:1$ to roughly $1:1$ implying a drastic rearrangement of orbital 
occupation (leading to para-orbital state?) across the I-M transition.  The
basic dependence of $J_{ij}^{ab}$ on orbital occupation and symmetry modifies
these as a consequence. Recovery of the corundum structure in an abrupt way.  

$(ii)$ Strong correlation driven physics as very clearly seen in optics 
and PES. Since $U,U',J_{H}$ are not likely to vary much across the MIT, 
the modification of hopping in a way consistent with (i) holds the key to 
increased itinerance. In any case, screening induced renormalisation of 
$U,~U'$ will occur only after the system has undergone an insulator-metal 
transition, and it is hard to understand how the transition itself can be 
``derived'' by reducing $U,U'$ in
the PI phase.

\subsubsection{In the AFM phase}

$(i)$ Non-FL features observed in transport studies in the AF-M phase, 
showing partial similarity to those observed in near-optimally doped 
cuprate superconductors raises interesting issues.  Is this one of the 
elusive examples of spin-charge separated metallic state in a 
three-dimensional oxide?

In a multi-orbital Mott-Hubbard scenario, strong coupling to coupled 
orbital-spin excitations should lead to a dynamically fluctuating hopping, 
leading to inhibition of AF-LRO and to strongly reduced coherence, 
manifested by a low Fermi temperature. The simultaneous observation of 
overdamped spin waves would also follow from such kind of effects.

\subsection{Implications for Theory}

A theoretical picture of the MIT in ${\rm V_{2}O_{3}}$ must address these 
issues 
in a consistent way.  Given that much more is known about the para-orbital,
paramagnetic state, as well as the view that understanding the AFI-AFM MIT 
requires a good knowledge of the PI, we focus our attention on the 
para-insulating/para-metallic states.  A complete understanding of the 
anomalous features near the AF-I/AF-M phase is beyond current theoretical 
capacity.

Castellani {\it et al.}~\cite{[8]} started with a single $c$-axis V-V pair 
in the real
crystal structure (RCS) of ${\rm V_{2}O_{3}}$, and solved the two-site cluster 
including $U,U'$ and $J_{H}$.  They assumed that screening processes reduce 
the values of these parameters, and, in particular, that $J_{H} \simeq 0.1U$.
With this choice, and in the situation where the $t_{2g}$ levels were split 
into an $a_{1g}$ singlet and $e_{g}^\pi$ doublet by the trigonal distortion, 
they found that the two electrons in the $a_{1g}$ orbitals on the pair form a 
total spin singlet, while the second electron populates the $E_{g}$ states.
The resulting model is clearly a $S=1/2$, two-orbital Hubbard model, with 
an orbital ordered, spin AF ground state.  Based on this picture, the one-band
Hubbard model was studied extensively for twenty years with a variety of
techniques.~\cite{[31]}  As is clear from the earlier discussion, a variety
of recent results run into direct conflict with the one-band modelling. 

Theoretically, the discrepancy has to do with the fact that $J_{H}$, which
controls the spin state at each V site, is very poorly screened in a solid.
This implies that $J_{H}$ in ${\rm V_{2}O_{3}}$ is larger than $0.1U$, the 
value used by Castellani {\it et al.}  Indeed, with $J_{H}>0.2U$,~\cite{[12]} 
the ground state has been found to have $S=1$, with a change to low-spin 
$S=0$ state as $J_{H}$ is reduced towards the value used by Castellani 
{\it et al.}~\cite{[8]}
 
  Given that the occupation of the $a_{1g},e_{g}^{\pi}$ orbitals changes 
discontinuously at the MIT, one would expect an important role for the 
trigonal field (since it acts like a fictitious external field in the orbital 
sector). This is expected to sensitively determine the occupancy of each 
orbital (orbital polarisation) in much the same way as the magnetisation of 
a paramagnet is a function of an applied magnetic field.  In particular, one 
expects that the lower-lying orbital(s) should be more localized in the 
solid ($e_{g}^{\pi}$), as we shall indeed find to be the case.  Further, 
the fact that the ratio of the orbital occupations changes discontinuously at
the MIT forces one to associate a corresponding change in the trigonal field 
as well. 

On the other hand, observation of global strong correlation signatures in 
various phases of ${\rm V_{2}O_{3}}$ as described above in detail implies 
a fundamental inadequacy of the band description, and mandates use of a 
strong correlation picture.  

Summarising, a consistent description of the PI/PM MIT requires a 
theoretically reliable description involving marriage of structural 
aspects (LDA) and strong correlation features (MO-DMFT). 

In the rest of this paper, we confine ourselves to the theoretical 
description of the PI/PM Mott transition in ${\rm V_{2}O_{3}}$.  Starting 
with a detailed exposition of the LDA+DMFT(IPT) which we use as a solver 
(the pros and cons of using IPT vis-a-vis other impurity solvers will be 
discussed), we will derive a two-fluid description of the PI/PM transition 
in ${\rm V_{2}O_{3}}$ attempting to achieve an internally consistent 
description. Finally, a quantitatively accurate description of the 
one-particle spectral function across the MIT {\it and} low-$T$ 
thermodynamics will be demonstrated within this scenario.   

\section{LDA+DMFT Technique}

As argued in detail and shown in recent work,~\cite{[23]} LDA+DMFT 
has turned out to be {\it the} method of choice for a consistent theoretical 
description of the competition between quasi-atomic, strong Coulomb 
interactions (multi-orbital) and itinerance (LDA spectra, encoding 
structural details in the one-electron picture) in real three dimensional 
transition-metal and rare-earth compounds.  The central difficulty in this 
regard has been the choice of an appropriate impurity solver to solve the 
multi-orbital, asymmetric Anderson impurity problem.  Two ways have been 
used with varying degrees of success: iterated perturbation theory (IPT) 
and quantum Monte Carlo (QMC).  

We have used multi-orbital extension of 
IPT to solve the impurity model.  On the one hand, such an approach 
should be valid if the behavior of the multi-orbital SIAM is {\it analytic} 
in $U,~U',~J_{H}$: this is known to hold for the general asymmetric version.  
MO-IPT also has the advantage of being extendable to $T=0$, and the 
self-energies can be extracted at modest numerical cost.  On the other hand, 
it is by no means exact, and calculations done for the one-band Hubbard 
model~\cite{[23]} show {\it quantitative} differences between IPT and QMC 
results for the critical value of $U=U_{c}$ at which the MIT occurs.  It 
has also been claimed~\cite{[24]} that the IPT spectral functions are very 
different from the QMC ones, and the latter are claimed to be more reliable 
vis-a-vis the {\it true} spectral function, as well as with the actual, 
experimentally determined spectral functions.  Here, we should emphasise 
that the IPT results for the many body DOS are in excellent agreement with 
{\it both} exact diagonalisation~\cite{[25]} as well as dynamical DMRG 
results for the one band Hubbard model in $d=\infty$. 
While no such evidence exists for multi-orbital models, we believe that 
the above arguments show that IPT is a good approximation, even though 
it is not ``numerically exact''.  

With these caveats, we describe our multi-orbital iterated perturbation 
theory (MO-IPT) for multi-band correlated systems.  For early TM oxides, 
one-electron bandstructure calculations show that, in three dimensional 
cases, the $t_{2g}$ DOS is well separated from the $e_{g}$ DOS as well as 
from the O-$2p$ DOS.  More precisely, the ``$t_{2g}$'' DOS does have 
contributions from components of the $e_{g}$ and O-$2p$ orbitals having 
$t_{2g}$ orbital symmetry.  Structural effects, such as those produced by 
trigonal crystal fields (${\rm V_{2}O_{3}}$~\cite{[34]}) and antiferroelectric 
distortions (${\rm VO_2}$~\cite{VO2}) are adequately described by LDA.  
In addition, the multi-orbital 
Coulomb interactions are parametrised by three parameters $U,U'$ and $J_{H}$.
The Hund's rule coupling, $J_{H}$, is very poorly screened and can be taken 
equal to its atomic value.  The intra-orbital ($U$) and inter-orbital ($U'$) 
Coulomb interactions are screened in the actual solid: usually, their 
screened values have traditionally been calculated using constrained 
LDA. In correlated systems, this is a problem, however, as the dynamical 
processes screening these parameters arise from {\it correlated} electrons 
having dualistic (itinerant-localised) character, rather than from free 
band electrons.  This well-known problem has received scant attention to 
date; indeed, we are aware of only one previous work attempting to cure this 
malady.~\cite{[26]} Below, we will show how the renormalised 
$U,~U'$ are self-consistently computed in a {\it correlated} approach, 
and lead to a consistent description of the PES results in the PM phase.

\subsection{The many-body Hamiltonian}
    
Generally, the full many-body Hamiltonian for early TMOs is written as,

\bn
\nonumber
H &=& \sum_{{\bf k} a b \sigma}(\epsilon_{{\bf k}a}
+\epsilon_{a}^{0}\delta_{ab})c_{{\bf k}a\sigma}^{\dag}c_{{\bf k}b\sigma} 
+ U\sum_{ia}n_{ia\uparrow}n_{ia\downarrow} \\ 
&+& U'\sum_{i a \ne b} n_{ia}n_{ib} 
- J_{H}\sum_{i a \ne b} {\bf S}_{ia}.{\bf S}_{ib}
\en
where $a,b=xy,yz,zx$ denote the three $t_{2g}$ orbitals. Details of the 
actual one-electron bandstructure in the real lattice structure are encoded 
in the one-electron band dispersion, $\epsilon_{{\bf k} a}$: the 
corresponding LDA DOS is 
$\rho(\omega)=N^{-1}\sum_{{\bf k} a}\delta(\omega-\epsilon_{{\bf k} a})$.  
Here $\epsilon_{a}^{0}=\epsilon_{a}-U(n_{a\sigma'}-\frac{1}{2})
+\frac{J_H}{2} \sigma (n_{a \sigma}-1)$,
where $\epsilon_{a}$ are the on-site energies of $t_{2g}$ orbitals within 
LDA and the rest of the terms are subtracted therefrom in order to avoid 
double-counting of interactions already treated on the average by LDA. 

\begin{figure}[h]
\epsfxsize=3.4in
\epsffile{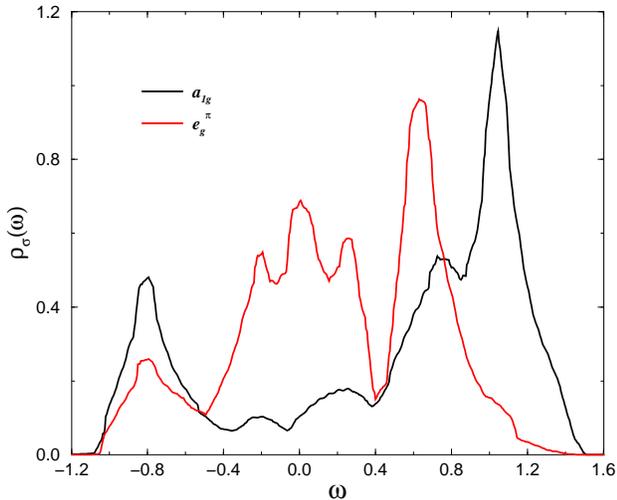}
\caption{
LDA partial density of states for the 
$e_{g}^{\pi}$ (red) and $a_{1g}$ (blue) orbitals, obtained from 
Ref.~\protect\onlinecite{Anisi}.} 
\label{fig0}
\end{figure} 

With these qualitative remarks, we now describe our multi-orbital LDA+DMFT 
procedure.  We emphasise that the basic method was already developed 
in Ref.~\cite{[23]}, and here, we extend this ideology using more 
detailed analysis to study the {\it full} one-electron spectral function 
in {\it both} insulating and metallic phases in ${\rm V_{2}O_{3}}$.  Our 
strategy is:

$(a)$ Beginning with LDA results in the real corundum lattice (see 
Fig.~\ref{fig0}), derive a correlated Mott insulating 
state using multi-orbital DMFT with $U=5~eV$, and $U'=3~eV$ (we use 
$J_{H}=1.0~eV$ for $V^{3+}$), values obtained from constrained LDA.
The LDA bandwidth is $W=2.5~eV$, and the bare LDA trigonal field is read 
off as $\Delta=0.32~eV$. In what follows, we will work in the basis of 
LDA eigenstates which diagonalises the one-particle density matrix.  

$(b)$ Mimic the effects of external pressure by noticing that it should 
lead to modification of the {\it renormalised} (correlated) value of the 
trigonal field.  In line with this ideology, first known to be propounded 
by Mott and co-workers, we search for the instability of the correlated 
(Mott insulator) solution found in (a), to a second solution of the DMFT
equations as a function of $\Delta$.  We emphasise here that we do {\it not} 
change the bare LDA parameters: indeed, we argue that a Mott transition from 
a correlated insulator to correlated metal {\it cannot} be validly described 
by changing bare LDA parameters, since these have no clear meaning in a 
strongly correlated system.

$(c)$ To provide a quantitative description of the one-electron spectral 
function in the metallic phase, we use the correlated (DMFT) results to 
compute the screening-induced reduction in $U,U'$ {\it in} the metallic 
phase.  This is crucial: we derive the screened $U,U'$ in the PM phase 
{\it after} deriving the I-M transition, and do not derive the I-M 
transition itself by reducing $U,U'$, as seemingly done in earlier 
work.~\cite{[23],Anisi} Using the DMFT result, the screened $U,~U'$ are 
estimated by an extension of Kanamori's $t$-matrix calculation to finite 
density.

Using the screened values of $U,U'$ (notice that $J_{H}$ is almost 
unaffected by screening, so we use the same value for it throughout), 
we compare our theoretical (correlated) DOS with PES and XAS results 
obtained experimentally in the PM phase.

Incorporation of electron correlations into the LDA gives rise to a 
two-stage renormalisation:

$(1)$ $U,~U'$ and $J_{H}$ give rise to multi-orbital Hartree shifts in 
the on-site orbital energies of each $t_{2g}$ orbital. In ${\rm V_{2}O_{3}}$, 
the trigonal field splits the $t_{2g}$ degeneracy, with the lowest 
$a_{1g}$ orbital [$\simeq (xy+yz+zx)$] lying about $\Delta=0.32~eV$ below 
the higher lying $e_{g}^{\pi}$ orbitals [$\simeq (xy-yz), (2xy-yz-zx)$] 
within LDA.  Given this, the $a_{1g}$ orbital is always occupied by one 
electron, the second residing in the $e_{g}^{\pi}$ orbitals.  The observation 
of $S=1$ on each V site requires strong $J_{H}$, implying even stronger 
$U,~U'$, even in the PM phase.  

Multi-orbital Hartree shifts renormalise the orbital energies: 
$\epsilon_{a_{1g}}=\epsilon_{0} + U'n_{e_g^\pi}$ 
($\epsilon_{e_g^\pi}=\epsilon_{0} + \Delta + U'n_{a_{1g}}$),
where $n_{e_g^\pi}$ ($n_{a_{1g}}$) is the $e_{g}^{\pi}$ ($a_{1g}$) 
orbital occupation.  These shifts 
correspond to effects captured by LDA+U.~\cite{[27]}  They do
give the correct, insulating ground states (with orbital/magnetic order), 
but {\it cannot} describe the phase transition(s) from correlated Mott 
insulators to correlated metals.  This can be traced back to the fact that 
LDA+U treats correlations on a static level, neglecting {\it quantum} nature
of electron dynamics, and so cannot access the spectral weight 
transfer-driven physics at the heart of Mott-Hubbard transitions.  
  
$(2)$ In a one-electron picture, this would be the end of the story. In 
reality, however, hopping of an electron from a given site to its 
neighbor(s) is accompanied by {\it dynamical} generation of particle-hole 
pairs (the more, the larger $U,U'$ are), which inhibit its free band motion.  
Electrons can move quasicoherently by dragging their corresponding 
``electronic polarisation cloud'' along. With increasing $U,~U'$, electrons 
get more and more ``localised'', corresponding to transfer of coherent low 
energy spectral weight to high-energy (quasi-atomic) incoherent regions, 
until at the MIT, {\it all} the weight resides in the incoherent Mott-Hubbard 
bands.  It is precisely this effect that is out of scope of LDA+U, and 
requires dynamical mean field theory (DMFT) for a consistent resolution.

Since the system is strongly correlated, the small changes in bare LDA 
parameters caused by $(1)$ lead to large changes in transfer of dynamical 
spectral weight.  Specifically, in systems undergoing MIT, small changes 
in bare LDA values of lattice distortion(s) transfer high-energy spectral 
weight to low energies, driving the Mott transition.

\subsection{The one-particle Green's functions}

 Given the actual LDA DOS for the $t_{2g}$ orbitals (this includes the 
V-$d$ orbitals and O $2p$ part having ``$t_{2g}$'' symmetry), the band 
Green's functions within the LDA (in the basis which diagonalises the 
one-particle density matrix) are 
$G_{ab}(\omega)=\delta_{ab}G_{a}(\omega)=\delta_{ab}\frac{1}{N} \sum_{\bf k}
(\omega-\epsilon_{{\bf k} a})^{-1}$. We define the correlated one-electron 
Green's function and the associated irreducible self-energy for each 
orbital $a$, by $G_{a \sigma}(\omega)$ and $\Sigma_{a\sigma}(\omega)$: 
the two are related by the usual Dyson's equation, 

\be
G_{a}^{-1}(\omega)=[G_{a}^0(\omega)]^{-1}-\Sigma_{a}(\omega) \;. 
\ee

It is obvious that the Green's functions can be exactly written down for the 
non-interacting case, as well as for the atomic limit 
($\epsilon_{{\bf k} a}=0$).  
In contrast to the case of the one-band Hubbard model, however, the 
exactly soluble atomic limit contains the local, inter-orbital correlation 
function, $<n_{a}n_{b}>$, in addition to $<n_{a}>$.  

\subsubsection{MO-IPT: an interpolative ansatz for multi-orbital systems}

In the spirit of the 
IPT developed by Rosenberg {\it et al.} for the one-orbital Hubbard case, 
we require an interpolative scheme that connects the two exactly soluble 
cases above, gives correlated Fermi liquid behavior in the metallic phase,
and a Mott-Hubbard transition from a correlated FL metal to a Mott insulator 
as a function of $U,U'$ for commensurate cases.~\cite{[28]}  In order 
to achieve this, we have extended the philosophy of Ref.~\onlinecite{[29]}.  
The central requirements for a consistent interpolative scheme capable of 
describing all of the above are that:

$(i)$ Formally defined one-electron Green's function, 

\be
G_{a}(\omega)=\frac{1}{N} \sum_{\bf k} \frac{1}{\omega + \mu
-\Sigma_{a}(\omega) -\epsilon_{{\bf k} a}}
\ee 
where $\epsilon_{{\bf k}a}$ describes the dispersion of the LDA bands for 
orbitals $a,b = t_{2g}$, and the self-energy is given by

\be
\Sigma_{a}(\omega)=\frac{\sum_{b}A_{ab}\Sigma_{ab}^{(2)}(\omega)}
{1-\sum_{b}B_{ab}\Sigma_{ab}^{(2)}(\omega)}
\ee
with 

\be
\Sigma_{ab}^{(2)}(\omega)=N_{ab}\frac{U_{ab}^{2}}{\beta^{2}}
\sum_{lm} G_a^0(i\omega_{l})G_b^0(i\omega_{m}) G_b^0(i\omega_{l}
+i\omega_{m}-i\omega)
\ee
being the second-order (in $U,U'$) contribution.  Here, $N_{ab}=2$ for 
$a,b=e_{g1}^{\pi},e_{g2}^{\pi}$ and $4$ for $a,b=a_{1g},e_{g1,2}^{\pi}$. 
Finally, the bath propagator is given as

\be
G_{a}^0(\omega)=\frac{1}{\omega+\mu_{a}-\Delta_{a}(\omega)} \;,
\ee 
with $\Delta_{a}(\omega)$ interpreted as the dynamical Weiss field for 
orbital $a$.

$(ii)$ The interpolative self-energy for each orbital $a$ should be chosen 
by fixing interpolative parameters such that the exact Friedel-Luttinger 
sum rule is strictly (numerically) obeyed, and,

$(iii)$ to reproduce the Mott insulator beyond a critical coupling, a 
high-energy expansion around the atomic limit is performed, yielding 
another equation for the interploative parameters.  Here, the high-energy 
expansion is truncated by including only the first few terms which 
guarantee the exact reproduction of the first three moments of the
one-electron spectral function.  In contrast to the one-band case, 
however, the exact atomic limit for the multi-orbital case contains the 
local, inter-orbital correlation function, 
$D_{ab}[n]=<n_{a}n_{b}>$.~\cite{Pou}  
We are aware of only one earlier work~\cite{[29]} where $D_{ab}[n]$ is 
computed using the coherent potential approximation (CPA).  Strictly 
speaking, this is an approximation to the Hubbard model(s) which is 
qualitatively valid in the Mott insulating state, but is known to fail 
in the correlated PM phase(s).  This is because CPA  replaces the actual, 
dynamical (annealed) ``disorder'' in the PM phase(s) by quenched, static 
disorder, and thus fails to capture the dynamical Kondo screening 
central to deriving correct (correlated) FL behaviour in the PM phase. Given 
this, it is hard to identify the extent to which computed results depend 
upon introducing such approximations, and this should be checked carefully 
by comparison with calculations which compute all local correlators in a 
single, consistent scheme. The correct way to compute $D_{ab}[n]$ is 
actually not complicated within multi-orbital IPT, and is described below.  

These two equations for the parameters $A_{ab}$ and $B_{ab}$ are solved 
to yield these as explicit functions of $U,U',<n_{a}>,<n_{a}n_{b}>$ and 
$<n_{a}^0>$ (this last average is the ``effective'' number of fermions in 
orbital $a$ corresponding to an ``effective'' Green's function used in the 
interpolative IPT, see Refs.~\onlinecite{[29],[30]}).  Explicitly, we have,

\be
A_{ab}=\frac{n_{a}(1-2n_{a})+D_{ab}[n]}{n_{a}^{0}(1-n_{a}^{0})}
\ee
and

\be
B_{ab}=\frac{(1-2n_{a})U_{ab}+\mu-\mu_{a}}{U_{ab}^{2}n_{a}^{0}(1-n_{a}^{0})}
\ee
where $n_{a}$ and $n_{a}^{0}$ are defined from the GFs $G_{a}(\omega)$ and 
$G_{a}^{0}(\omega)$.  The inter-orbital correlation function $D_{ab}[n]$ is 
calculated from

\be
D_{ab}[n] = <n_{a}><n_{b}> + \frac{1}{U_{ab}} 
\int_{-\infty}^{+\infty} \Xi(\omega)
f(\omega) 
d\omega \;, 
\ee
the last term following from the equation of motion for $G_{a}(\omega)$ and 
$\Xi(\omega) \equiv -\frac{1}{\pi} {\rm Im} 
\left[ \Sigma_{a}(\omega)G_{a}(\omega) \right]$.
 
The above equations form a closed set of coupled, non-linear equations 
which are solved numerically. We found fast convergence of the 
self-consistent system of equations, and typically twenty iterations sufficed 
for the parmeter values considered here.  The converged results allow us 
to study the one-particle DOS, and the corresponding orbital occupations, 
spin states, as well as the strength and character of local multi-orbital 
correlations in {\it both} PI and PM phases, as described below in detail. 

\section{Results and discussion}

In this section, we present a detailed set of results for the one-particle 
spectral function in {\it both} the PI and PM phases in ${\rm V_2O_3}$. While 
doing so, we will make extensive contact and discuss important differences 
between our work here and previous results recently obtained by other 
authors.~\cite{[23],Anisi} 

\begin{figure}[htb]
\epsfxsize=3.2in
\epsffile{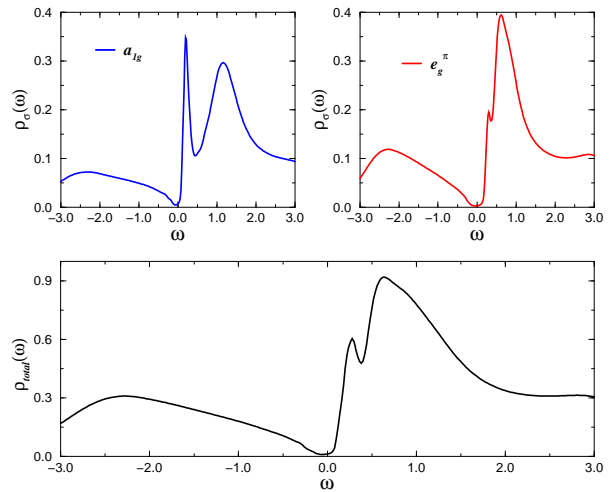}
\caption{
Orbital-resolved (upper panels) and total (lower
panel) one electron spectral function for the insulating phase of 
${\rm V_2O_3}$ obtained with $5.0$~eV and $J_H=1.0$~eV.} 
\label{fig1}
\end{figure}

In Fig.~\ref{fig1}, we show the single particle DOS for the Mott insulator, 
obtained with $U=5.0~eV, U'=3.0~eV$ and $J_{H}=1.0~eV$ as correlation 
parameters for this system.  These are slightly different from those used 
in our previous work,~\cite{[34]} but are roughly the same as those used 
by Held {\it et al.} recently.~\cite{Anisi} A clear Mott-Hubbard gap, 
$E_{G}=0.2~eV$, is seen, and, as expected from the orbital assignment, 
the $e_{g}^{\pi}$ states are more localised in the solid. The renormalised 
trigonal field $\Delta_{t}^{r}=\delta_{a_{1g}} -\delta_{e_g^\pi}=0.32~eV$, 
is read off directly from Fig.~\ref{fig1}. The orbital occupations are 
computed to be ($n_{a_{1g}},n_{e_{g1}^\pi},n_{e_{g2}^\pi}=0.32,0.34.0.34$) 
in the PI, in nice agreement with XAS estimations. 

We now study the paramagnetic metallic state obtained as an instability of 
the correlated Mott insulator derived above.  In other works, the PM state 
is ``derived'' not by searching for an instability of the correlated PI 
state under pressure, but by computing the LDA bandstructure for the 
``metallic'' state {\it without} correlations.  The screened values of 
$U,~U'$ are computed using constrained LDA, and these are then used to 
describe the PM phase.  In reality, however, one has to study the transition 
to the PM phase without leaving the correlated picture, and derive the 
transition by searching for the second, metallic solution of the DMFT 
equations under pressure.  

To justify our new approach, we specify the problems associated with 
earlier approaches:~\cite{[23],Anisi,Bier}

$(1)$ It is theoretically inconsistent to derive a phase transition between 
two strongly correlated phases by using corresponding LDA bandstructures 
to separately derive the two phases.  This is because using changes in bare 
LDA parameters to study correlated phases is clearly problematic, since 
these parameters have no clear meaning in a correlated picture.  One must 
use the {\it renormalised} values of these parameters instead, and these 
are generically modified in unknown ways by strong multi-orbital 
correlations.  These changes in bare LDA parameters, and the modification 
of the response of correlated electrons to these changes, 
must be selfconsistently derived {\it within} the LDA+DMFT procedure.
Clearly, this route has not been used in other approaches.

$(2)$ It follows that an inescapable consequence of using such approaches 
is that the values of $U,U'$ used for the PM phase are computed using 
constrained LDA (i.e, using the uncorrelated bandstructure assuming that 
the screening electrons are {\it free} band electrons).  However, in 
reality, the screening electrons in the correlated PM phase have a dualistic 
character generic to the Mott-Hubbard character of the system.  As is 
known from Ref.~\cite{[31]}, the electronic kinetic energy, or itinerance, 
is reduced in the PM phase: it is these correlated electrons  which 
screen $U,~U'$ in the real correlated system.  In an ``ab initio'' treatment,
the effective $U,U'$ should be computed using the {\it correlated} spectral 
functions to estimate screening.  Replacement of the renormalised spectral 
functions by bare LDA ones will introduce an approximation, overestimating 
the screening of $U,~U'$ (this is hard to quantify, but is estimated to 
be of order of twenty percent!).
 
In order to avoid these difficulties, we adopt the following strategy.  

$(A)$ We hypothesise that external pressure modifies the trigonal field.  
To our knowledge, this is not completely new: Mott and co-workers proposed 
such ideas in the seventies,~\cite{[1]} and more recently, Tanaka made 
a similar hypothesis in a cluster approach for ${\rm V_2O_3}$.~\cite{[11]}
To model this change in $\Delta_{t}$ under pressure, we {\it do not} 
change the trigonal field by hand.  Rather, we input trial values of 
$\Delta_{t}$, changing it from its value in the (Mott) PI by small 
trial amounts, and {\it search} numerically for its critical value, 
$\Delta_{t}^{c}$, which stabilises the second, correlated metallic 
solution of the DMFT equations.  The new values of $\Delta_{t}$ in the 
(correlated) PM phase are again read off from the converged DOS for each 
orbital.  We emphasise that we do not decrease $U,U'$ by hand, neither 
do we use different LDA DOS for different phases, for reasons explained 
before.  We note that Savrasov et al.~\cite{[32]} have employed similar 
ideology to study the giant volume collapse across the $\alpha-\delta$ 
transition in $Pu$.

In a strongly correlated system, small changes in the {\it renormalised} 
trigonal field lead to large changes in dynamical spectral weight transfer 
from high- to low energies, typically over a scale of a few $eV$.  This 
is precisely our mechanism for the first-order Mott transition in 
${\rm V_2O_3}$ under pressure.  We expect the free energy to have a double 
well structure.  Pressure changes the trigonal field (we remind the reader 
that $\Delta_{t}$ acts like an external field in the orbital sector), 
lowering the second minimum (PM) relative to the first (PI) beyond 
$\Delta_{t}^{c}$. 

$(B)$ Using the converged DOS for each orbital, the occupation(s) of 
various orbitals (and their changes from their PI values), the local 
spin value at each V site, and information about the detailed character 
of the PM state is directly obtained.   

\begin{figure}[htb]
\epsfxsize=3.2in
\epsffile{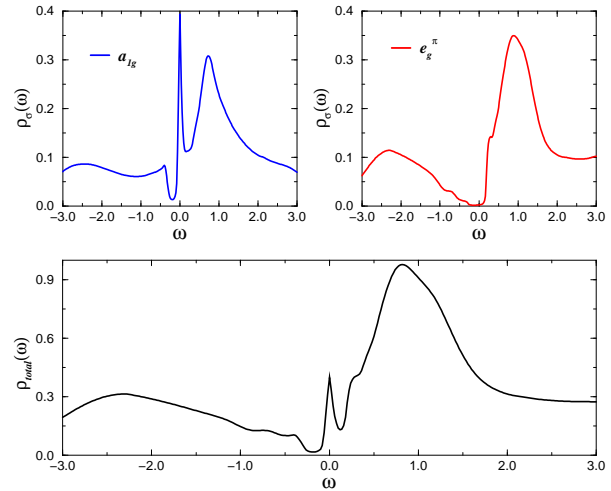}
\caption{(color online) Orbital-resolved (upper panels) and total 
(lower panel) one electron spectral functions for the metallic phase of 
${\rm V_2O_3}$.  Note that only the $a_{1g}$ orbital DOS crosses $E_{F}$
in the metallic phase; the $e_g^\pi$ orbitals still shows Mott-Hubbard 
insulating features, showing the ``two-fluid'' character of the MIT in 
${\rm V_2O_3}$.} 
\label{fig2}
\end{figure}

In Fig.~\ref{fig2}, we show our results for the PM phase obtained within 
our technique.  At $T=0$, the hypothetical PM phase (it is never observed 
in reality) shows a sharp, quasicoherent FL resonance.  We identify this 
feature with combined spin-orbital Kondo screening in the PM phase of a 
multi-orbital Hubbard model. This is easily seen as follows.  To obtain 
a correlated Mott insulator, we need not only $U=5.0$~eV, but also 
$U'\simeq (U-2J_{H})=3.0$~eV: indeed, if $U'$ were ignored, a $t_{2g}$ 
electron hopping from one V site to its neighbor could always hop off 
like a band electron just by going into an unoccupied $t_{2g}$ orbital 
at that site, making a PI state impossible.  Given the small spectral 
weight carried by this feature, we expect a low lattice coherence scale, 
above which the PM would be described as an incoherent metal. The trigonal 
field in the PM, $\Delta_{t}'=-0.291$~eV and the occupations of each 
$t_{2g}$ orbitals, $[n_{a1g},n_{eg1},n_{eg2}=0.38,0.31,0.31]$, are 
read off from the converged PM solution of the DMF equations.  Very 
satisfyingly, the spin state remains unchanged, and the orbital occupations 
change across the MIT in semiquantitative agreement with XAS 
results.~\cite{[10]}  

\begin{figure}[htb]
\epsfxsize=3.2in
\epsffile{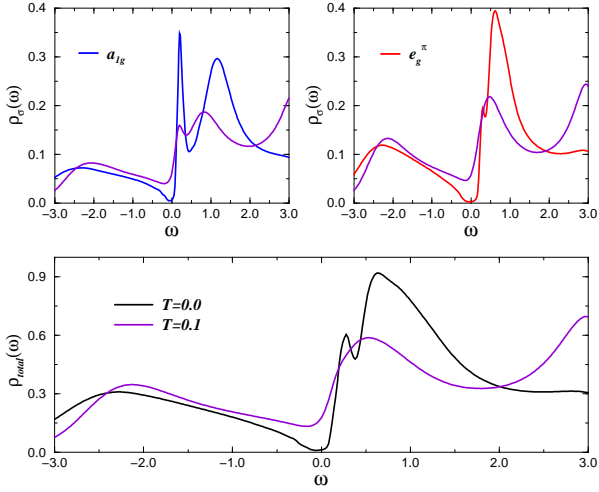}
\caption{
Effect of temperature $(T)$ on the orbital-resolved 
(upper panel) and total (lower panel) one electron spectral functions 
for the insulating phase of ${\rm V_2O_3}$.} 
\label{fig3}
\end{figure}

\begin{figure}[htb]
\epsfxsize=3.2in
\epsffile{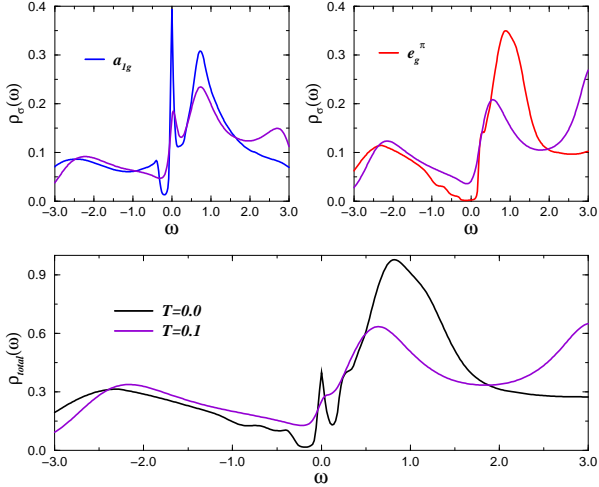}
\caption{
Effect of temperature $(T)$ on the orbital-resolved 
(upper panels) and total (lower panel) one electron spectral functions 
for the metallic phase of ${\rm V_2O_3}$.} 
\label{fig4}
\end{figure}

\begin{figure}[htb]
\epsfxsize=3.2in
\epsffile{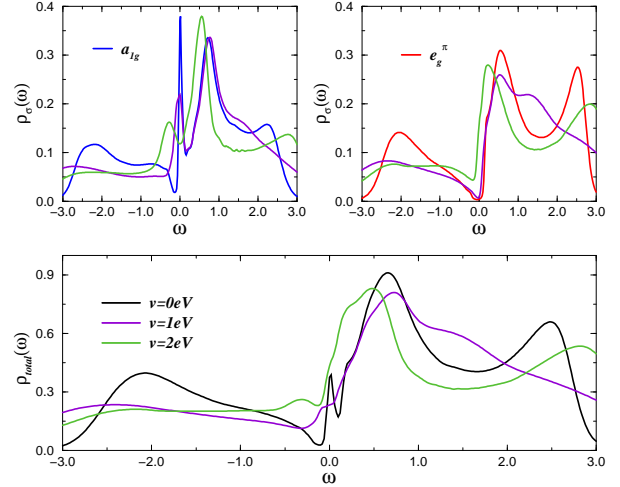}
\caption{
Effect of disorder $(v)$ on the orbital-resolved 
(upper panels) and total (lower panel) one electron spectral functions 
for the metallic phase of ${\rm V_2O_3}$.} 
\label{fig5}
\end{figure}

In Figs.~\ref{fig3} and~\ref{fig4}, we show the effect of finite temperature 
on our results.  As expected on general grounds within the DMFT framework, 
the FL resonance is broadened by finite-$T$ and lowered in height (the 
pinning of the interacting DOS at $E_{F}$ to its LDA value, dictated by 
Luttinger's theorem, holds only at $T=0$). The effects of introducing 
chemical disorder in the PI is shown in Fig.~\ref{fig5}.  The results 
were obtained by combining multi-orbital IPT with the coherent-potential 
approximation (CPA)~\cite{[33]}.  In agreement with very recent 
observations,~\cite{Mo-2004} we indeed observe a broadened ``quasiparticle'' 
in the PM, and closing in of the Mott gap in the PI by incoherent spectral 
weight transferred across large energy scales from high- to low energies.  
Comparing Figs.~\ref{fig3} and~\ref{fig5}, it is clear that, at sufficiently 
high-$T$, the spectra in the chemically disordered PI and the PM phases do 
resemble each other qualitatively.  As observed by Allen,~\cite{JAllen} 
this implies that there is no fundamental difference between the ``metal'' 
and ``insulator'' at sufficiently high $T$: this agrees with the observation 
that the first order Mott transition is replaced by a smooth crossover at 
high $T$.

An extremely important conclusion follows directly from an examination 
of the orbital-resolved DOS in the PM phase.  We find that the 
$e_{g}^{\pi}$ orbital DOS shows ``Mott insulating'' (see Fig.~\ref{fig2}) 
behavior, while only the $a_{1g}$ orbital DOS is responsible for the 
metallicity.  This constitutes an explicit realisation of the ``two-fluid'' 
model used phenomenologically in connection with the MIT in disordered 
semiconductorsin the past.~\cite{[1]}  In Refs.~\cite{[34],VO2} we 
already showed the orbital selective character, as well as the evolution of  
the DOS at $E_{F}$ as a function of the occupation of the $a_{1g}$ orbital.  
A clear first-order I-M transition around $n_{a_{1g}}=0.38$ was found, 
involving, as described above, a discontinuous change in (selfconsistently 
determined) occupations of each orbital. These observations are intimately 
linked to the multi-orbital Mott-Hubbard character of correlations in
$V_{2}O_{3}$.  Polarised XAS results might already hold the clue to 
establishing an approximate two-fluid character of the PM phase:  the 
$a_{1g}$ spectral weight should dominate over the $e_{g}^{\pi}$ 
contribution for energies up to the Mott gap.  Orbital resolved optical 
studies could also be used to test our picture.  
 
Strong indirect support for our picture comes from the early 
observation~\cite{[9]} of an anisotropic change in the lattice 
constants along $a/b$ (planar) and $c$ axes across the P-MIT in 
${\rm V_2O_3}$. Instead of a uniform volume collapse expected across the 
MIT,~\cite{[35]} increase in $a(b)$ and a {\it decrease} in $c$ was 
found across the MIT.  Such an anisotropic volume change across the MIT 
is inconsistent with {\it simultaneous} gapping of all $t_{2g}$ orbitals
(where we would expect an isotropic volume change),
but is completely consistent with our (orbital selective) two-fluid picture
derived above.

\section{COMPARISON WITH PES AND XAS}

In this section, we describe how our approach provides an excellent 
description of the experimental photoemission (PES) and X-ray absorption 
(XAS) data on ${\rm V_2O_3}$ in the PM phase.  As argued before, this requires 
us to recompute the full one-particle local spectral function (total DOS) 
using values of $U,U'$ renormalised by dynamical metallic screening in the 
{\it correlated} metallic phase. In order to do this, we have used an 
extension of Kanamori's $t$-matrix approach~\cite{[3]} to estimate 
these parameters.  

In the multi-orbital case, this is a horrendous problem in general.  
Fortunately, in the effective two-fluid picture of the PM phase derived 
above, the general analysis can be simplified. This is because the 
$e_{g}^{\pi}$ electrons remain ``insulating'', i.e, Mott localised, up to 
energies of the order of the Mott-Hubbard gap.  We then expect only the 
$a_{1g}$ electrons to provide efficient screening, and so consider only 
the $a_{1g}$ band in the computation of the effective $U,~U'$ below.  
In general, we need the full ${\bf q}$-dependent particle-particle 
susceptibility for this purpose.  Using the LDA+DMFT Green function for 
the $a_{1g}$ orbital, 

\be
\chi_{pp}({\bf q})=-\frac{1}{N} \sum_{n m {\bf k}}
G_{a_{1g}}({\bf q}-{\bf k},i\omega_{m}-i\nu_{n})
G_{a_{1g}}({\bf k},i\nu_{n}) \;.  
\ee
In $d=\infty$, this can be 
expressed as an integral over the LDA DOS and the full irreducible 
one-electron self-energy~\cite{[31]}, permitting a direct evaluation. 
The onsite Hubbard $U$ is renormalised by the local part of this 
susceptibility, via the equation,

\be
U_{eff}=\frac{U}{1+U\chi_{loc}^{'}(\omega)} \;.
\ee

Using the relation $U\simeq (U'+2J_{H})$, valid for $t_{2g}$ systems, 
along with the fact that $J_{H}$ is essentially unscreened, we estimate 
$U,~U'$ in the PM phase. We observe that this implies a frequency-dependent 
$U_{eff}=U(\omega)$. We have found, however, on computation that the 
$\omega$-dependence is weak for energies up to the Mott gap, and so use 
its $\omega=0$ value $U_{eff}=U(0)$ in what follows.

\begin{figure}[htb]
\epsfxsize=3.4in
\epsffile{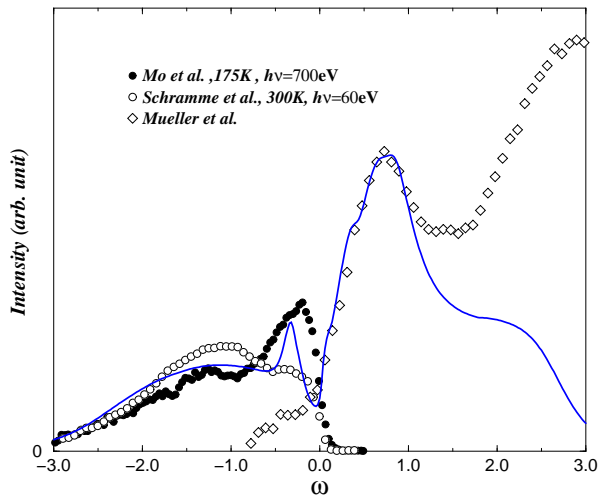}
\caption{(color online) 
Comparison of theoretical LDA +DMFT result (blue) for the total one-electron 
spectral function in the metallic phase of ${\rm V_2O_3}$ to the experimental 
results taken from Refs.~\protect\onlinecite{[20],[21]} (for PES) and from 
Ref.~\protect\onlinecite{[IPES]} (for XAS).} 
\label{fig6}
\end{figure}

Starting with the values of $U,~U'$ used earlier, we estimate 
$\chi_{loc}^{'} (0) \simeq 0.084$, yielding $U_{eff} \simeq 3.5$~eV.  
With $J_{H}=1.0~eV$, this implies that $U'\simeq 1.5$~eV.  We 
have recomputed the one-electron spectral function for the PM phase using 
these values.  The results are compared with experimental 
work~\cite{[20],[21],[IPES]} in Fig~\ref{fig6}.  Very satisfyingly, excellent 
quantitative agreement over almost the whole energy scale from 
$-3.0\le\omega\le 1.2$~eV is clearly observed.  In addition to the detailed 
shape of the lower Hubbard band (in PES), excellent agreement with the 
intense peak in XAS is also clear.  Consideration of parts of the spectrum 
for $\omega\le -3.0$~eV and $\omega\ge 1.2$~eV is hampered by our 
restriction to the $t_{2g}$ sector in the LDA+DMFT calculations. Due to the 
reduction of $U,U'$ as above, the $t_{2g}$ orbital occupation is now
$(n_{a1g},n_{eg1},n_{eg2})=(0.36,0.32,0.32)$, in even better agreement with
XAS results.~\cite{[10]}

However, though good, the agreement is not quite so perfect in the 
low-energy region: our computed``'broad'' peak (ascribed to a 
``quasiparticle'' in earlier work~\cite{[21]}) is narrower than the 
experimental feature by a factor of 1.8.  On first sight this might seem 
to confirm the interpretation in the earlier work.  However, we observe 
that this feature is peaked at $\omega=-0.37$~eV, while a clear 
pseudogap-like dip is resolved around $E_{F}$.  Hence, in our picture, the 
metallic phase cannot be described in a FL quasiparticle language; instead,
short-lived, incoherent, non-FL pseudoparticles should dominate the PM phase.
Interestingly, observation of a linear-in-$T$(instead of the $T^{2}$ form 
for a correlated FL) resistivity supports a non-FL quasiparticle interpetation.
It is possible that the $T$ regime where this is valid lies {\it above} an 
effective FL coherence scale (below which a $T^{2}$ term in resistivity would
follow) which is masked by emergence of orbital/spin ordered insulating states
at lower $T$.  At $T>T_{coh}$, the $dc$ resistivity is indeed linear in $T$ 
in a Hubbard model framework, where it arises from inelastic scattering off
unquenched spin-orbital local moments in a $d=\infty$ multiband Hubbard model.

Our observation of a low energy pseudogap feature can be traced back to the 
strong bonding-antibonding splitting observed in LDA results (see 
Fig.~\ref{fig0}). This is a direct consequence of strong hopping along 
the $a_{1g}$ orbitals, leading to strong covalency and robust singlet 
character between V-V pairs along the $c$-axis. (see Ref.~\onlinecite{[36]} 
for an early discussion on this point).  Our analysis does
partially show up the effects of strong covalency, manifested in the pseudogap
feature found above.  However, the fact that we can resolve most of the 
spectrum accurately, but fail to reproduce the correct broadening of the 
low-energy feature, implies that it may be necessary to explicitly consider 
the dynamical effects of intersite (V-V) correlations for a complete 
resolution of the PES spectrum, as alluded to in the experimental section.  
This is however out of scope of LDA+DMFT, and requires a cluster extension.

\section{Conclusion}

  In conclusion, we have studied the first-order Mott transition under 
pressure in ${\rm V_{2}O_{3}}$ using the state-of-the-art LDA+DMFT technique.
We have proposed a new picture for the MIT, which is driven by large changes
in the transfer of dynamical spectral weight (via DMFT) accompanying small
changes in the renormalised trigonal field splitting under pressure.  Very 
good quantitative agreement with the orbital occupations, spin state of 
$V^{3+}$ ions, as well as effective mass enhancement in the PM state is 
obtained.  The MIT is found to be first-order, and orbital selective (only
the $a_{1g}$ DOS shows metallic behavior).
Finally, using the correlated solution, we have computed the
screening induced renormalisation of $U,U'$ in the PM phase.  Using these,
excellent quantitative agreement with the full one-particle spectral function
(PES {\it and} XAS) is found in the PM phase.  These findings constitute 
strong support for our underlying two-fluid picture, which is ultimately an 
interesting manifestation of strong, multi-orbital Coulomb interactions in 
this early transition-metal oxide.    

\acknowledgments
The authors would like to acknowledge H. Tjeng for valuable discussions.
The work of LC was carried out under the auspices of the 
Sonderforschungsbereich 608 of the Deutsche Forschungsgemeinschaft.
MSL acknowledges financial support from the EPSRC (UK).

\end{document}